\documentclass[%
 aip,
 amsmath,amssymb,
 reprint,%
]{revtex4-1}

\usepackage{graphicx}
\usepackage{dcolumn}
\usepackage{bm}
\usepackage{fancyhdr}

\usepackage[utf8]{inputenc}
\usepackage[T1]{fontenc}
\usepackage{mathptmx}
\usepackage{mathtools}
\usepackage{romannum}
\usepackage{xcolor}
\usepackage{setspace}

\begin{document}

\preprint{AIP/123-QED}

\title[]{Quantum sensing with superconducting circuits}

\author{S. Danilin}
\email{sergey.danilin@glasgow.ac.uk.} 
\author{M. Weides}
\affiliation{James Watt School of Engineering, University of Glasgow, Glasgow G12 8QQ,
	 United Kingdom}

\date{\today}

\begin{abstract}
Sensing and metrology play an important role in fundamental science and applications, by fulfilling the ever-present need for more precise data sets, and by allowing to make more reliable conclusions on the validity of theoretical models. Sensors are ubiquitous, they are used in applications across a diverse range of fields including gravity imaging, geology, navigation, security, timekeeping, spectroscopy, chemistry, magnetometry, healthcare, and medicine. Current progress in quantum technologies inevitably triggers the exploration of quantum systems to be used as sensors with new and improved capabilities. This perspective initially provides a brief review of existing and tested quantum sensing systems, before discussing future possible directions of superconducting quantum circuits use for sensing and metrology: superconducting sensors including many entangled qubits and schemes employing Quantum Error Correction. The perspective also lists future research directions that could be of great value beyond quantum sensing, e.g. for applications in quantum computation and simulation. 
\end{abstract}

\maketitle

\section{\label{sec:Introduction}Introduction}

Quantum sensing is the procedure of measuring an unknown quantity of an observable using a quantum object as a probe. Quantum objects \textemdash\ those in which quantum-mechanical effects can manifest and be observed \textemdash\ are known to be highly sensitive to even tiny changes in their environment which is inevitably coupled to them. These changes can be so small that it is extremely challenging, or even impossible, to detect them employing classical measurements. Consequently, the fact that the probe/sensor is quantum endows it with extreme sensitivity. A back-action imposes a random change of a system state during the measurement. A probability of an outcome depends not only on the initial state of the system but also on the strength of the measurement~\cite{Back-action_general}. In the case of a quantum sensor, the back-action is quantum-limited, and measurement schemes where it can be evaded have been demonstrated, e.g. in Ref~[\onlinecite{Back-action_evading}].

Quantum sensors are highly engineered systems for measurements ranging from gravitational pull, to magnetic and electric fields and propagating photons. Different quantum systems have been employed for sensing to date, we are giving a short overview in the following.

Thermal vapors of alkali atoms closed in a cell, pumped, and interrogated by near-resonant light are used to measure magnetic fields~\cite{Budker_OpticalMagnetometry}. This method is also known as nonlinear magneto-optical rotation magnetometry. Magnetometers of this type do not have intrinsic $1/f$-noise due to the absence of nearly degenerate energy states and do not require cryogenic cooling for operation; they offer millimetre spatial resolution and sensitivity exceeding ${\rm fT}/\sqrt{\rm Hz}\ $ \cite{Kominis_subfemtotelsa}. Their accuracy is shot-noise limited and scales as~\cite{Budker_magnetometry_review,Allred_SERF_magnetometry} $\delta B\sim1/\sqrt{NT_{2}t}$, where $N$ is the number of atoms, $T_{2}$ is the transverse relaxation (dephasing) time, and $t$ is the time of the signal acquisition. Spin-exchange relaxation free (SERF) operation can be achieved by increasing the gas density, and improves the sensitivity of atomic magnetometers~\cite{Allred_SERF_magnetometry}. Another type of magnetic field sensor utilises ensembles of nuclear spins~\cite{Waters_magnetometer_NMR}. Although they are not as sensitive as atomic vapor sensors, they find applications in a variety of areas from archaeology to MRI systems~\cite{Degen_sensor_review} due to simplicity and robustness. 

Nitrogen vacancy centres (NV-centres) in diamond \textemdash electron spin defects \textemdash have recently attracted a lot of attention as quantum sensors, with predicted sensitivity for ensembles of spins $\sim0.25\cdot {\rm fT}/\sqrt{{\rm Hz}\cdot {\rm cm}^3}$~\cite{Taylor_diamond_magnetometer}, and experimentally achieved sensitivities of $\sim 1 {\rm pT}/\sqrt{\rm Hz}$~\cite{Wolf_NVensemble}. With the advent of single spin in diamond readout~\cite{Gruber_single_defect_spectroscopy,Dobrovitski_single_spin_control}, it became possible to use such single spins for magnetometry~\cite{Taylor_diamond_magnetometer,Balasubramanian_nanoscale_imaging_magnetometry,Cole_decoherence_microscopy}, sensing of electric fields~\cite{Dolde_electric_field_sensing_NV_spin}, or to measure pressure~\cite{Doherty_NV_centre_pressure_sensor}. Demonstrations of frequency standards based on the NV defect centres in diamond~\cite{Hodges_NV_centre_frequency_standard} and nanoscale thermometry with down to $5 {\rm mK}/\sqrt{\rm Hz}$ sensitivity have been made~\cite{Kucsko_NV_centre_thermomtry,Neumann_NV_centre_thermomtry,Toyli_NV_centre_thermometry}. The main advantages of this type of sensors are their stability in nanostructures and superior $10-100$ nm spacial resolution.

Trapped ions have been employed to detect extremely small forces and displacements. To increase the solid angle of the field access to the trapped ion, an enhanced access ion trap geometry was shown~\cite{Maiwald_enhanced_access_ion_trap}. A force sensitivity of $\sim 100 {\rm yN}/\sqrt{\rm Hz}$ has been reached on the crystals of trapped atomic ions, with the ability to discriminate ion displacements of $\sim 18 {\rm nm}$~\cite{Biercuk_trapped_ions_force_detection}. Their augmented force and displacement sensitivity are often traded against the reduced resolution. Rydberg atoms are another physical system for quantum sensing of electrical fields. Their high sensitivity is based on huge dipole moments of highly excited electronic states~\cite{Osterwalder_Rydberg_states_sensing}. Rubidium atoms prepared in circular Rydberg states were used for non-destructive (quantum nondemolition~\cite{Braginsky_QND_measurements,Levenson_QND_optics,Braginsky_QND,Grangier_QND_meaurement_optics}) measurement of single microwave photons~\cite{Nogues_QND_photon_measurement,Gleyzes_quantum_jumps_of_light}, and sensitivities reaching $3 {\rm mV}/{\rm m}\sqrt{\rm Hz}$ were achieved when Schr{\"o}dinger-cat states~\cite{Hacker_NaturePhotonics_cat_states_Rb,Vlastakis_100Photon_cat_state} were involved in the protocol~\cite{Facon_SchrodingerCat_electrometer}. 
The reader is directed to reviews on atomic spectroscopy and interferometry based sensors~\cite{Kitching_review}, quantum metrology with single spins in diamond~\cite{Chen_spins_in_diamond_review}, comparative analysis of magnetic field sensors~\cite{Lenz_magnetic_sensors_review}, and a more general and comprehensive review on quantum sensing~\cite{Degen_sensor_review}. 

Superconducting quantum circuits are among the leading approaches to real-world applications with quantum computers due to their controllability and reproducibility. Here we review their past and explore their future use as quantum sensors.
\begin{figure*}[!t]
	\includegraphics[width=\textwidth]{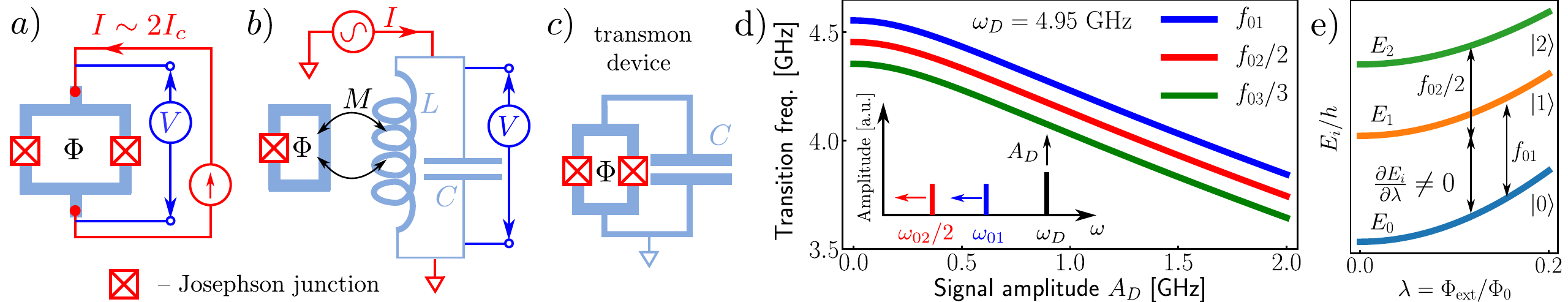}
	\caption{Principal measurement schemes for magnetometers based on Superconducting Quantum Interference Devices (SQUIDs) for dc- (a) and rf- (b) measurement strategies. (c) The circuit of a frequency-tunable transmon device. (d) ac Stark shifts of transmon qudit transition frequencies induced by an external microwave drive of frequency $\omega_D$. The shifts depend on the frequency detuning of the signal from the transitions and the amplitude $A_D$ of the signal. (e) Dependence of the energy spectrum of a tunable transmon device on the normalized external magnetic flux $\lambda=\Phi_{\rm ext}/\Phi_0$.}
	\label{figure_1}
\end{figure*}
The long-established Superconducting Quantum Interference Devices (SQUIDs) for sensing of magnetic fields are distinguished from the relatively new quantum sensors based on superconducting qubits. Due to the noise, the accuracy of the measured parameter usually scales with sensing time as $\sim1/\sqrt{t}$, known as the Standard Quantum Limit (SQL) scaling. The ultimate accuracy scaling law, known as the Heisenberg Limit (HL), is given by the uncertainty principle and improves better as $\sim1/t$. The figures of merit of any sensor are: i) its accuracy limit, ii) the time it takes to reach this limit, and iii) the dynamic range of values possible to measure. With this in mind, this perspective aims to draw attention to, and trigger the development and experimental testing of, sensing protocols, which will allow to improve upon the SQL in the time domain without loss in sensor dynamic range.

The article is structured as follows: Section \Romannum{2} briefly describes magnetometers based on Superconducting Quantum Interference Devices; sensors employing superconducting qubits are discussed in Section \Romannum{3}; the utility of quantum entanglement for sensing is considered in Section \Romannum{3}.1, and prospects to use Quantum Error Correction are discussed in Section \Romannum{3}.2; finally, Section \Romannum{4} summarises the conclusions of this perspective and provides an outlook on future experiments.

\section{\label{sec:SQUIDs}Superconducting Quantum Interference Devices}

Shortly after the theoretical prediction of the Josephson effect~\cite{Josephson_superconductive_tunneling} and its experimental observation~\cite{Anderson_tunneling_current_exp}, the quantum interference of currents was demonstrated. This interference is at the core of any SQUID magnetometer operation~\cite{Jaclevic_currents_interference}. There are two types of SQUID-based magnetometers: dc-SQUID~(Fig.\ref{figure_1}(a)) with a pair of Josephson junctions connected in parallel in a superconducting loop, and rf-SQUID~(Fig.\ref{figure_1}(b)) with a single junction in a loop. The experimental methods and measurement schemes of magnetic field sensing with SQUID systems are diverse and extensively studied~\cite{Clarke_handbook,Fagaly_SQUID_review}. SQUIDs became the most sensitive tools for magnetic field measurements, with applications in geophysics and neuroscience~\cite{Clarke_highTc_SQUIDs}, for example, and record magnetic flux sensitivities of $\sim 50\ {\rm n}\Phi_0/\sqrt{\rm Hz}$ ($\sim 50\ {\rm nT}/\sqrt{\rm Hz}$) at $100\ {\rm Hz}$ and $\sim 50\ {\rm nm}$ loop diameter ~\cite{Vasyukov_nanoSQUID}. SQUID superiority in sensitivity has only recently been challenged with the advent of SERF atomic vapor magnetometers~\cite{Allred_SERF_magnetometry}. Despite SQUIDs high sensitivity, the accuracy of measured results at low frequencies is shot-noise limited, and improves as $\sim1/\sqrt{t}$~\cite{Clarke_handbook}.

\section{\label{sec:Qubits}Superconducting Quantum Circuit Based Sensors}

Superconducting circuits including macroscopic, human-designed, many-level anharmonic systems (qubits/qudits Fig.\ref{figure_1}(c)) are a well established experimental technology platform in the field of quantum computation and simulation. The field development gained significant momentum when limitations of the conventional classical paradigm of computation became apparent in the early 1980s~\cite{Feynman_simulations_computer}. At present, it is undergoing a transition to the so-called Noisy Intermediate-Scale Quantum~\cite{NISQ2018_Preskill} regime, and new applications of superconducting circuits comprising qubits/qudits in quantum sensing and metrology are emerging. 

The first experimental works where such circuits are used as quantum sensors have recently appeared. The frequency and amplitude of a microwave signal were determined by spectroscopic means~\cite{Schneider_spectroscopy_sensing} and with time-domain measurements~\cite{Kristen_time-domain_sensing} via ac Stark shifts of qudit higher energy levels (Fig.\ref{figure_1}(d)). Here, an external microwave signal, with a frequency $\omega_D$ and an amplitude $A_D$, shifts the transitions from their unperturbed values. The change in transition frequencies allows for a measurement of $A_D$ and $\omega_D$ for the applied signal to be made. Furthermore, the absolute power flowing along a transmission line~\cite{Honigl-Decrinis_power_sensor} or distortions of microwave control pulses ~\cite{Bylander_distortion_sensing,Gustavsson_distortion_sensing} were measured by strong coupling to a flux qubit. Methods to use a transmon qubit as a VNA for in situ characterization of the transfer function of xy-control lines~\cite{Jerger_PRL}, and as a cryoscope to compensate the distortions of z-control pulses~\cite{Rol_APL} were demonstrated recently. These methods are useful for the calibration of microwave lines and the deduction of power reaching the circuit at millikelvin temperatures. They allow for the correction of pulse imperfections and increase fidelities of control gates used in quantum computation and simulation. All of these methods are implemented on a superconducting structure comprising a single qubit/qudit.

In quantum information processing, precise dynamic control of the quantum states is key to increasing the circuit depth. Conventional qubit frequency tuning is achieved by applying a well-controlled magnetic flux through a split junction loop within the quantum circuit. In turn, the quantum circuit can sense these externally generated static or dynamic fields. By replacing the flux-threaded split junction with a voltage biased junction (gatemon~\cite{Larsen_PRL,Lange_PRL}), the sensed external quantity becomes a voltage instead of a current. Note that we are using magnetic flux as an external parameter here, but the sensed quantity could be a voltage too. Superconducting circuits comprising qubits/qudits possess all the properties required to construct external field sensing quantum systems~\cite{Degen_sensor_review}: they have quantized energy levels; it is possible to initialize, coherently control, and read out their quantum states; and energy levels of the circuit, $E_i(\lambda)$, can be made dependent on the external parameter, $\lambda$, to be measured (Fig.~\ref{figure_1}(e)). For frequency-tunable qubits with a split junction, the parameter $\lambda$ is an external flux, $\Phi_{\rm ext}$. If the qubit is prepared in a superposition of basis states $\{0,1\}$ and placed in an external field, its state will accumulate phase $\phi(\Phi_{\rm ext}) = \Delta\omega(\Phi_{\rm ext})\cdot\tau$, dependent on the flux $\Phi_{\rm ext}$. $\Delta\omega(\Phi_{\rm ext})=\omega_q(\Phi_{\rm ext})-\omega_{\rm d}$ is the detuning between the qubit and the control pulse frequency used for the state preparation. By applying a second control pulse identical to the first one after some time $\tau$, and measuring the population of qubit basis states, it is possible to reveal the accumulated phase in oscillating dependencies of $P_{|0\rangle}$ and $P_{|1\rangle}$. This measurement, known as {\it Ramsey fringes interferometry}, can be employed for field sensing tasks. An equal superposition state $(|0\rangle+|1\rangle)/\sqrt{2}$  provides the maximal pattern visibility here, and the best sensitivity to the field.

\begin{figure}[!ht]
	\includegraphics[width=\linewidth]{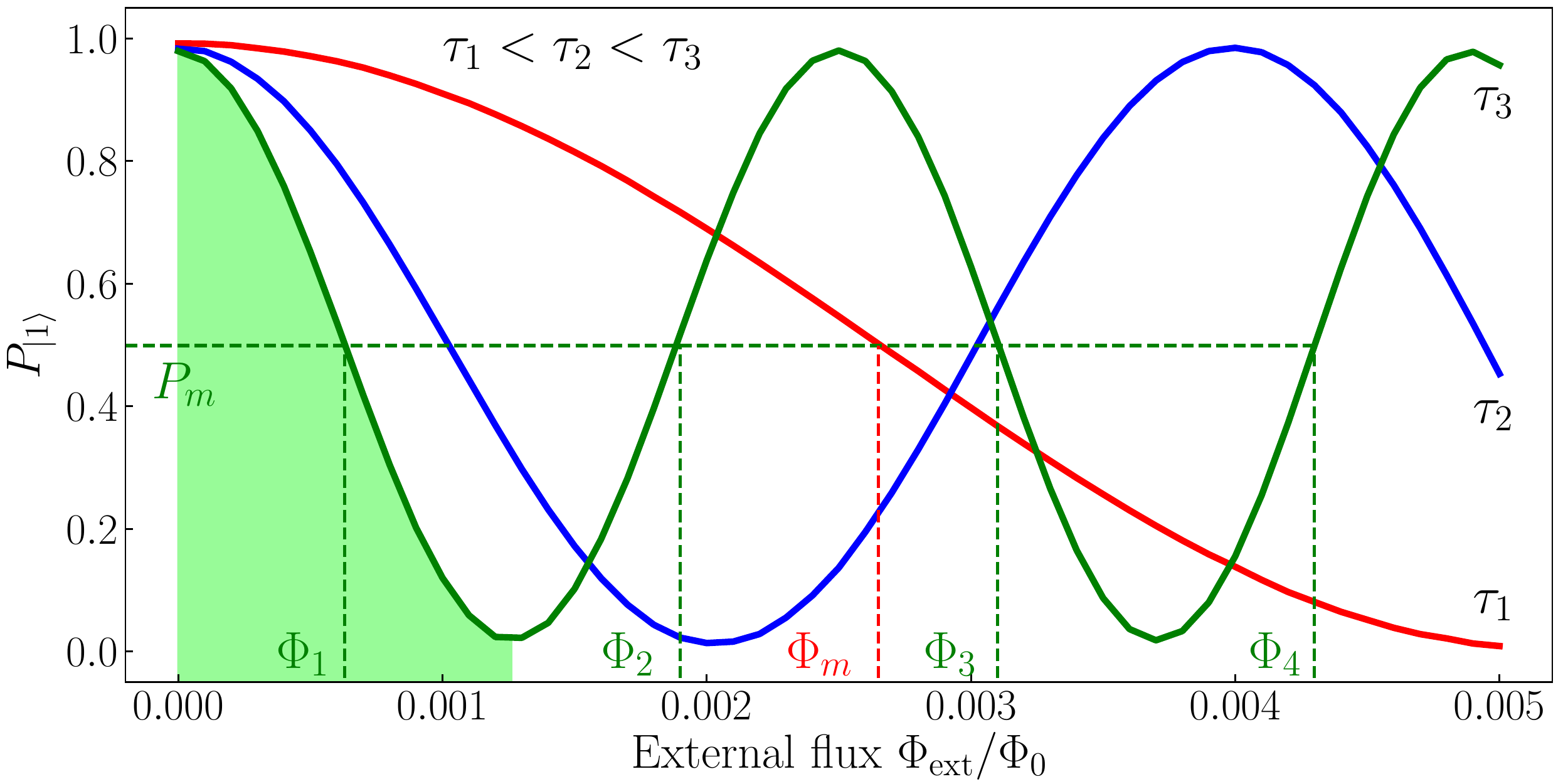}
	\caption{Ramsey fringes pattern $P_{|1\rangle}(\Phi_{\rm ext},\tau)$ at different delay times $\tau_i$. $P_m$ is an outcome measured during the sensing procedure and used for the determination of the unknown flux value.}
	\label{figure_3}
\end{figure}

The Ramsey fringes pattern $P_{|1\rangle}(\Phi_{\rm ext},\tau)$ can be simulated or directly measured as a calibration pattern before the field sensing routine. In this scenario, the outcome $P_m$ measured during the sensing procedure will be used in conjunction with the calibration pattern to determine the unknown flux value. Fig.~\ref{figure_3} shows the simulated dependence of the probability $P_{|1\rangle}(\Phi_{\rm ext},\tau)$ on the external flux at different delay times $\tau_i$. One can see that the longer the delay time, the higher the sensitivity of $P_{|1\rangle}$ to the external flux. This is only the case if delay time is shorter than the coherence time $T_2^*$ of the qubit; for longer delay times, the sensitivity will be reduced. However, two issues should be noted here. Firstly, for an unknown flux value, it is not possible to choose {\it a priori} the delay time $\tau^*$ with the best sensitivity. Secondly, for longer delay times, it is not possible to unambiguously determine the measured flux based on a single outcome. As shown in Fig.~\ref{figure_3}, the same result $P_m$ can correspond to many flux values $\{\Phi_1,\Phi_2,\Phi_3,\Phi_4\}$, and to make the measurement unambiguous, one has to reduce the dynamic range of the sensor to the interval highlighted in the figure. This interval is substantially shorter than for the shortest delay time $\tau_1$, where the measurement is single-valued (one-to-one correspondence).

The phase estimation algorithms~\cite{Giovannetti_1,Giovannetti_4}(PEA) can be employed to address both issues. They gradually tune the delay time to the value $\tau^*=T_2^*$ with the highest available sensitivity, without a reduction in the dynamic range of the sensor, and appear to be powerful tools in sensing. Kitaev~\cite{Kitaev} and Fourier~\cite{Quantum_Fourier} phase estimation algorithms were used with a single tunable transmon qubit to measure external flux, and experimentally demonstrated the accuracy scaling beyond the SQL~\cite{Danilin_magnetometry}. These algorithms involve a stepped strategy. At each step of the Kitaev algorithm, the interval of possible fluxes is reduced by a factor of two based on the measurement outcome, and a new optimal delay time is found for the next step providing improved sensitivity. The optimal delay times grow from step to step on average, and gradually tend to the coherence time. PEAs allow to approach $\sim1/t$ accuracy scaling (HL). The qubit coherence $T_2^*$ serves as a quantum resource. The longer it is, the more steps of the algorithm can be made before the delay time approaches $T_2^*$, and higher accuracy can be achieved in the same sensing time. Quantum sensing algorithms employing qutrits instead of qubits have  also recently been considered~\cite{Shlyakhov_metrology}.

\subsubsection{Adding Entanglement}

\begin{figure}[!ht]
	\includegraphics[width=\linewidth]{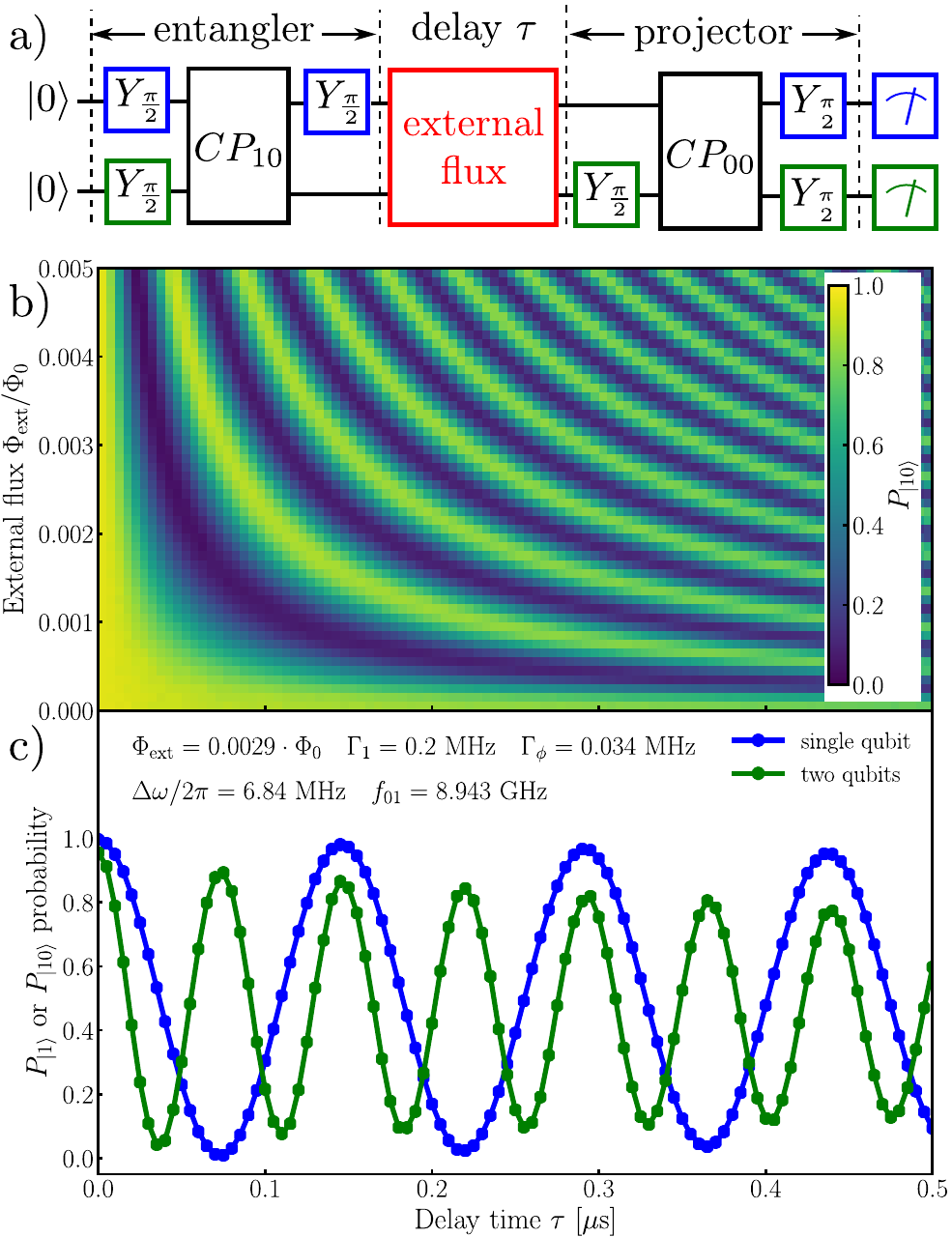}
	\caption{(a) Sequence of control operations used in the simulation. (b) Simulated probability pattern $P_{|10\rangle}(\Phi_{\rm ext},\tau)$ of the two-qubit state $|10\rangle$ with $\Gamma_1=0.2\ {\rm MHz}$, $\Gamma_\phi=0.034\ {\rm MHz}$, and $f_{01}=8.943\ {\rm GHz}$ for both qubits. (c) Comparison of time dependencies of $P_{|1\rangle}$, probability for a single qubit, and $P_{|10\rangle}$, probability for two-qubit case, at a fixed external flux.}
	\label{figure_2}
\end{figure}

Quantum entanglement can provide improvements in attainable sensitivity~\cite{Giovannetti_2,Giovannetti_3} for short interrogation times $\tau$ since entangled state of $N$ qubits, used as the probe state, allows for an $N$-fold speed-up in phase accumulation. Experimentally, this has been demonstrated for systems including three trapped $^9Be^+$ ions~\cite{Leibfried_Science}, four-entangled photons~\cite{Nagata_Science}, ten nuclear spins~\cite{Jones_Science}, or single bosonic mode of a superconducting resonator~\cite{Wang_NatCommun}. Though these experiments clearly demonstrate the improvement of sensitivity beyond the SQL with the number of entangled qubits, they do not yet provide an explicit metrologic routine to allow the measurement of an unknown external field. To this end, we analyze the probability pattern $P_{|10\rangle}$ of a two-qubit state for sensing with PEAs. 
We use entangling conditional phase gates~\cite{DiCarlo_cPhase} and simulate the evolution of the two-qubit state in QuTiP~\cite{QuTiP}. Fig.~\ref{figure_2}(a) shows a time scheme used for the simulation of the pattern, where $CP_{ij}$ denotes the c-Phase gate inverting the sign of only the $|ij\rangle$ state. Relaxation $\sqrt{\Gamma_1}\hat\sigma_+$ and pure dephasing $\sqrt{\Gamma_\phi/2}\hat\sigma_z$ processes were taken into account in the simulation with the identical value used for both qubits decoherence rates. The flux dependence of the qubit transition frequency is assumed to be the same for both qubits. They are equally detuned to the flux point where $d\Delta\omega/d\Phi_{\rm ext}\ne 0$. Starting from both qubits in the ground state, we create the $|\Phi^+\rangle$ Bell state, apply the external flux we want to measure to both qubits, and allow the system to evolve for a variable time $\tau$. After that, we convert the entangled state to a separable state, projecting the entangled state phase to the phase of the first qubit, shown in Eq.~(\ref{evolution}). 

\begin{multline}
|00\rangle\xrightarrow[CP_{10}]{\rm Entangler}\frac{|00\rangle+|11\rangle}{\sqrt{2}}\underset{\tau}{\Longrightarrow}\frac{|00\rangle+e^{i\phi(\Phi_{\rm ext},\tau)}|11\rangle}{\sqrt{2}}\\
\xrightarrow[CP_{00}]{\rm Projector}\left(\frac{-1+e^{i\phi}}{2}|0\rangle+\frac{-1-e^{i\phi}}{2}|1\rangle\right)\otimes|0\rangle.
\label{evolution}
\end{multline}
Subsequent measurement of both qubit states, for different delay times $\tau$ and different external fluxes $\Phi_{\rm ext}$, results in a pattern $P_{|10\rangle}$, shown in the Fig.~\ref{figure_2}(b), and allows for the determination of the probabilities of all four possible two-qubit states. The pattern closely resembles that of the {\it Ramsey fringes}, but has doube the frequency of $P_{|10\rangle}$ oscillations, $\phi=2\times\Delta\omega\times\tau$ (Fig.~\ref{figure_2}(c)). 

The doubling of phase accumulation speed results in two times better accuracy of flux sensing at the same short sensing times. However, the pattern contrast also reduces more quickly (Fig.~\ref{figure_2}(c)) in comparison
with a single qubit case, making the advantage less impressive for long measurement times. The quicker reduction of pattern contrast originates from the shortening of the coherence time $T_{2,N}^*$ with the growth of the system size $N$. It was pointed out~\cite{Huelga_PRL} that for entanglement only pure dephasing provides the same maximal sensitivity reached at a shorter delay time $\tau$, in comparison with the standard {\it Ramsey fringes} scheme, because the coherence time shortens proportionally to the size of the system ($\sim N$). However, experimental tests on up to eight trapped-ion qubits, under the influence of correlated noise~\cite{Monz_PRL}, demonstrated a quicker coherence reduction ($\sim N^2$). So, the pure dephasing rate can be proportional to $\sim N^\alpha$, with $\alpha=1$ representing non-correlated noise and $\alpha=2$ representing correlated noise acting on all qubits. Experimental investigations into noise correlations between two or more  superconducting qubits have only recently started to appear~\cite{Harper_NatPhys,Lupke_PRX,Han_FundRes}. These results are important for quantum computation and quantum-enhanced sensing, and further experiments would be of great value. The dependence of the entangled state coherence time on the number of entangled qubits useful for quantum sensing has not been studied for superconducting qubits thus far.

\begin{figure}[!th]
    \includegraphics[width=\linewidth]{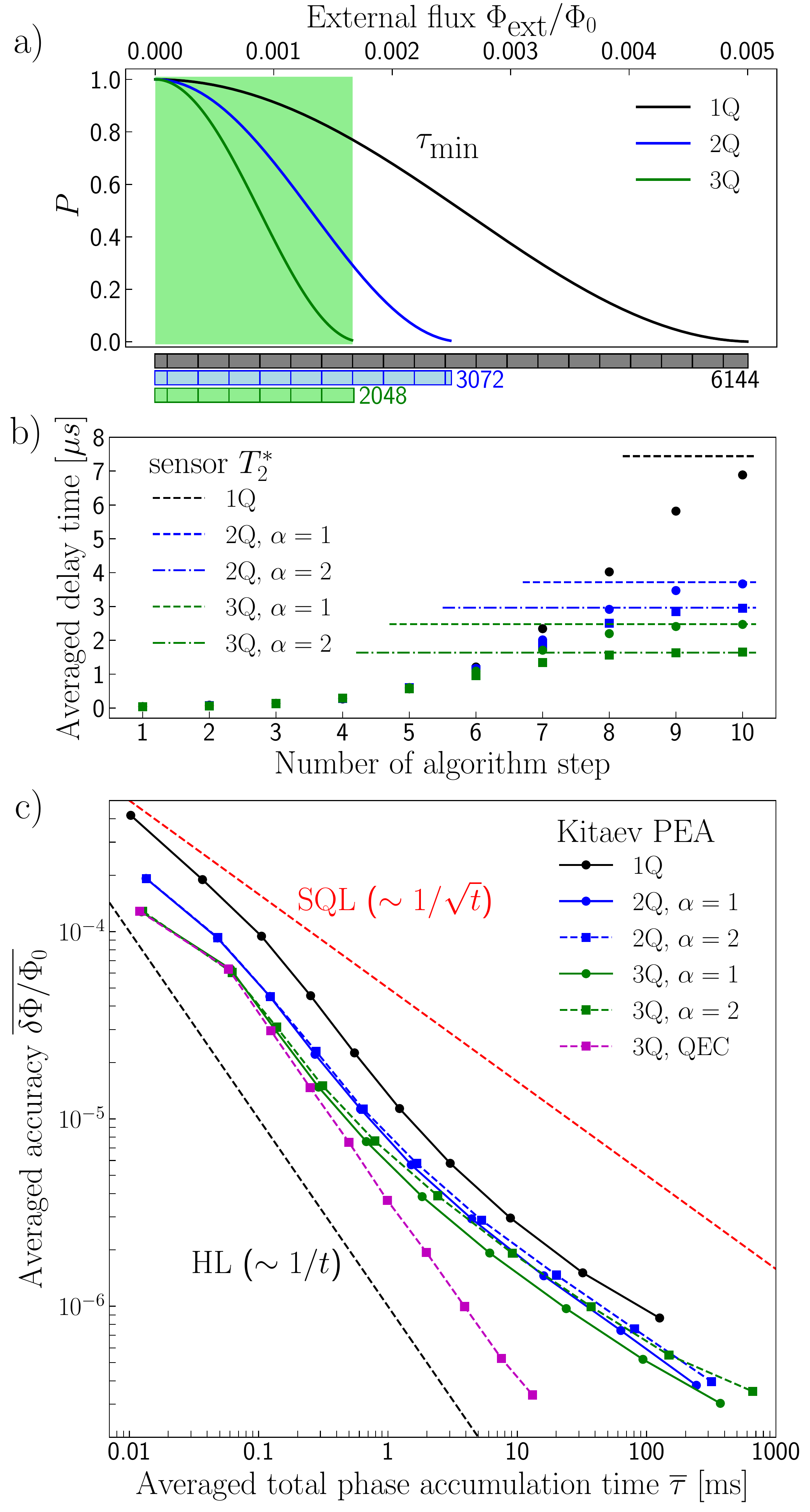}
    \caption{(a) Calibration patterns at the minimal delay time used in the simulations for the single-qubit sensor, and for the sensors with 2 and 3 entangled qubits. (b) Averaged delay times for different steps of the Kitaev PEA and sensors comprising the different number of qubits. (c) Comparison of the flux sensing accuracy scaling with the phase accumulation time for the sensors comprising the different number of qubits. The coherence time $T_2^*$ is set to infinity for the QEC case.}
    \label{figure_4}
\end{figure}

Next, we simulate the flux sensing routines based on the Kitaev PEA run with a single qubit, and with two and three qubits prepared in the GHZ entangled state. We compare the accuracy of flux sensing achieved by employing entangled states to that of a single qubit, for the cases when 
$\alpha = 1\ \textrm{and}\ 2$ in the pure dephasing rates. To perform the simulation we compute the probability
patterns $P_{|10...0\rangle}$ of the $N$-qubit states for $N = 1,2,\ \textrm{and}\ 3$ as
\begin{equation}
    P_{|10...0\rangle}(N,\Delta\omega,\tau)=\frac{1}{2}+\frac{1}{2}e^{-\left(\frac{N\Gamma_1}{2}+N^\alpha\Gamma_\phi\right)\tau}\cos(N\Delta\omega\tau).
    \label{P_pattern}
\end{equation}
These probabilities are obtained after projecting the phase accumulated by the GHZ $N$-qubit state during the evolution in the external magnetic field to the first qubit.
The dependencies of the qubits' spectra on the flux are assumed to be identical. The total relaxation rate and the total pure dephasing rate are $\Gamma_{1,N}=N\Gamma_1$ and $\Gamma_{\phi,N}=N^\alpha\Gamma_\phi$, with $\Gamma_1 = 0.2\ \textrm{MHz}$ and $\Gamma_\phi = 0.034\ \textrm{MHz}$. We use the equidistant flux grids in the computation of probability patterns with 2048, 3072, and 6144 values for 1-, 2-, and 3-qubit cases, respectively. If the sensor is exposed to the measured field only during the phase accumulation time, the dynamic range of fluxes measured with $N$ entangled qubits is $\Delta\Phi_{\rm ext}\sim\pi/N\tau_{\rm min}$, where $\tau_{\rm min}$ is the minimal time required for switching the external field on and off. Thus, for a sensor with $N$ entangled qubits, the dynamic range is reduced as $\sim1/N$ in comparison with a single qubit sensor. The flux grids for 2- and 3-qubit sensors form the subsets of the grid for the single-qubit sensor (Fig.\ref{figure_4}(a)). We choose $F=256$ flux values to be measured from the flux grid of the 3-qubit sensor so that it is also possible to measure them with the two other sensors. As the flux interval of possible values is reduced by 2 at each step of the Kitaev PEA, the chosen flux grids allow us to make 10 steps of the algorithm. We repeat the algorithm $M=24$ times at each of the $F=256$ flux values. Fig.~\ref{figure_4}(b) shows the obtained delay times for every step of the algorithm averaged, first, over all $M=24$ repetitions and, then, over all $F=256$ flux values. One can see that the delay times grow on average from step to step, and tend toward the coherence time of the sensor $T_2^*$. With the reduction of the coherence time for $N=2,\ 3$ or $\alpha$ going from 1 to 2 the delay times start to saturate at the earlier steps. 

Fig.~\ref{figure_4}(c) shows the results of the simulations. We compute the phase accumulation time $\tau_{j,k,l}$ for every flux value $(j)$, repetition $(k)$, and the step $(l)$, and then the averaged total phase accumulation time, $\overline{\tau_l}$, for every step as
\begin{equation}
\overline{\tau_l}=\frac{1}{F}\sum_{j=1}^F\frac{1}{M}\sum_{k=1}^M\tau_{j,k,l},\quad \tau_{j,k,l}=\sum_{i=1}^l\tau_i^{(j,k)}n_i^{(j,k)}.
\label{phase_acc_time_eq}
\end{equation}
Here, $\tau_i^{(j,k)}$ and $n_i^{(j,k)}$ are the delay time for the step number $i$ and the number of measurements done at this step for the $j$-th flux value in the $k$-th repetition, respectively.
In our simulations, we use $\sigma_0=\sigma_1=1.5$ for the widths of measurement outcome normal distributions for states $|0\rangle$ and $|1\rangle$, and  $\epsilon=0.01\%$ for the error probability. These determine the number of measurements $n_i$ done at each step, and also the condition to terminate the step and discard less probable flux values. By the end of step $l$ of the algorithm, we have a probability distribution for the remaining most probable fluxes for every flux value $\Phi_j, j\in[1,F]$ chosen to be measured, and every repetition $k\in[1,M]$. We use this distribution to compute the mean flux values $\hat\Phi_{jkl}$ and find the averaged flux accuracy for every step as
\begin{equation}
\overline{\left(\frac{\delta\Phi}{\Phi_0}\right)_l}=\sqrt{\frac{1}{\Phi_0^2F}\sum_{j=1}^F\frac{1}{M-1}\sum_{k=1}^M(\hat\Phi_{jkl}-\Phi_j)^2}.
\label{accuracy_eq}
\end{equation}
Dependencies of the averaged flux accuracy $\overline{\left(\delta\Phi/\Phi_0\right)_l}$ on the averaged total phase accumulation time $\overline{\tau_l}$ are shown in Fig.~\ref{figure_4}(c).

We compare the improvements of the flux sensing accuracy with the phase accumulation time for the sensors comprising a single qubit, or 2 and 3 entangled qubits with $\alpha=1$ or $2$, in Fig.~\ref{figure_4}. One can see that for the first algorithm steps, the scaling of the flux accuracy is close to the HL scaling for all considered sensors. When the averaged delay time approaches the coherence time and starts to saturate (Fig.~\ref{figure_4}(b)), the accuracy scaling deviates from the HL scaling and returns gradually back to the SQL scaling. The shorter the coherence time of the sensor, the sooner this transition happens, so that the sensors with 2 and 3 entangled qubits deviate from the HL scaling at the earlier steps of the algorithm. Nevertheless, the accuracies at the same phase accumulation time achieved by the sensors with the entangled qubits are always better than that of the sensor based on a single qubit, and the sensor with 3 entangled qubits proves to be better than with 2 entangled qubits. The advantage in the accuracy is reduced as the crossover from the HL scaling to the SQL scaling occurs, but the advantage from the earlier steps of the algorithm is not completely lost at the later steps even when $\alpha>1$. Importantly, for all pure dephasing rates with $\alpha\in[1,2]$, there is an accuracy improvement caused by the use of the entangled sensor. 

In practice, the calibration of a sensor employing PEA -- the measurement of the probability pattern $P_{|10...0\rangle}$ -- can take a long time. To mitigate this, FPGA-based electronics can be used for fast reset of the sensor qubits~\cite{Gebauer_AIP}. 
If the duration of control pulses and the time to read out and reset the qubits are much shorter than the coherence time of the sensor, the total sensing time will almost entirely consist of the phase accumulation time. This will noticeably shorten the calibration and speed up the sensing itself. 

Another experimental aspect of employing entangled states for sensing is the sensitivity of the control pulses to the external field being measured. Conditional phase gates realized via flux control pulses are very sensitive to external magnetic fields, which makes it necessary to allow the external field to act on the system only during the phase accumulation time. Otherwise, the initial entangled $N$-qubit state will not be the desired one. With regard to this, all-microwave entangling gates~\cite{all_microwave_gates} can be considered as an alternative way of the preparation of the sensing state. If they appear to be more resilient to the field being measured, it will be possible to keep the field continuously present, simplifying the operation. 

\subsubsection{Quantum Error Correction}

Instead of increasing the phase accumulation speed in the system of many entangled sensors, one can try to improve the coherence time $T_2^*$ of the available sensor qubits. This can be achieved with Quantum Error Correction (QEC) strategies~\cite{QEC_theor_1,QEC_theor_2} already demonstrated with superconducting qubits~\cite{QEC_exp_1,QEC_exp_2,QEC_exp_3}. In this case, one has a set of sensor qubits and entangles them with ancillary qubits which are periodically measured to detect any possible errors. Entanglement of a sensor-ancillary pair allows a judgment on the state of the sensor to be made, based on the measurement of the ancillary. Once an error is detected in one of the ancillary qubits, the state of the corresponding sensor qubit is corrected. It was shown that for $d-$dimensional sensor space it is sufficient to have the same dimensionality of the ancillary space, and that a minimalistic two-qubit QEC setup (one sensor qubit and one ancillary qubit) can outperform in resolution schemes involving large-scale entanglement~\cite{Sekatski_theory}. This strategy is experimentally very attractive due to its relative simplicity and the fact that it has not yet been tried with superconducting qubits.

In the limiting case of infinitely long coherence time $T_2^*$ of sensor qubits achieved with QEC, the accuracy scaling will follow the HL law for all steps of the algorithm (Fig.~\ref{figure_4}(c)). The number of steps and the final accuracy reached will depend on the number of flux values used in the calibration pattern.

Recently, important results were obtained on the attainability of the HL scaling of precision in time, depending on the noise present. For a superconducting qubit sensor with the Hamiltonian $\omega_q\hat{\sigma}_z/2$, it is possible to use QEC to compensate "perpendicular" $\{\hat{\sigma}_x,\hat{\sigma}_y\}$ noise (relaxation) and achieve the HL precision scaling. "Parallel" $\{\hat{\sigma}_z\}$ noise (pure dephasing) can not be compensated by QEC, and SQL scaling remains unsurpassed~\cite{HL_in_metrology}. As a consequence, superconducting qubits with coherence time limited by the relaxation rate are more suitable for sensing tasks. The design of a superconducting circuit and the metrologic scheme including QEC for sensing have already been suggested to overcome the limit imposed by relaxation~\cite{QEC_against_relaxation}.

It is important to emphasize that QEC schemes rely on the realization of fast and full quantum control, where qubit readout, analysis, and reaction times are much shorter than the coherence time of the sensor. The development of experiments where PEAs are combined with QEC is of great importance, as such experiments can contribute substantially to future progress of quantum sensing and metrology.

\section{\label{sec:Conclusions}Conclusions}

Two strategies involving entanglement for sensing and metrology with superconducting quantum circuits are considered, with the aim of going beyond the SQL scaling in the time domain. The first is based on the increased speed of phase accumulation for a large-scale entangled state of $N$ sensors. The advantages seen for this strategy depend on the characteristics of noise seen by the entangled state. Future experimental studies of the coherence reduction with the size of the entangled state $N$ are interesting and necessary, but have not yet been undertaken with superconducting qubits. The second strategy is to use QEC on entangled pairs (sensor-ancillary) of superconducting qubits, with the idea of enhancing the coherence time of the sensor qubits. To this end, implementation of metrological protocol combining QEC with one of the PEAs could experimentally demonstrate magnetic field sensing beyond the SQL scaling in time. Proposed experiments are of high value for quantum metrology and sensing. 

\begin{acknowledgments}
The authors are thankful to Prof A.V. Lebedev for valuable discussions and to P.G. Baity and J. Brennan for the careful reading of the manuscript. They acknowledge the financial support from the EPSRC grant number EP/T018984/1. 
\end{acknowledgments}

\ 

{\bf\noindent Data availability}

The data that support the findings of this study are available from the corresponding author upon reasonable request.

\bibliography{quantum_sensing_perspective}

\begin{thebibliography}{10}

\bibitem{Back-action_general}
M.~Hatridge, S.~Shankar, M.~Mirrahimi, F.~Schackert, K.~Geerlings, T.~Brecht,
  K.~M. Sliwa, B.~Abdo, L.~Frunzio, S.~M. Girvin, R.~J. Schoelkopf, and M.~H.
  Devoret.
\newblock {\em Science}, 339:178, 2013.

\bibitem{Back-action_evading}
J.~B. Hertzberg, T.~Rocheleau, T.~Ndukum, M.~Savva, A.~A. Clerk, and K.~C.
  Schwab.
\newblock {\em Nature Physics}, 6:213, 2010.

\bibitem{Budker_OpticalMagnetometry}
D.~Budker and M.~Romalis.
\newblock {\em Nature Physics}, 3:227, 2007.

\bibitem{Kominis_subfemtotelsa}
I.~K. Kominis, T.~W. Kornack, J.~C. Allred, and M.~V. Romalis.
\newblock {\em Nature}, 422:596, 2003.

\bibitem{Budker_magnetometry_review}
D.~Budker, W.~Gawlik, D.~F. Kimball, S.~M. Rochester, V.~V. Yashchuk, and
  A.~Weis.
\newblock {\em Rev. Mod. Phys.}, 74:1153, 2002.

\bibitem{Allred_SERF_magnetometry}
J.~C. Allred, R.~N. Lyman, T.W. Kornack, and M.V. Romalis.
\newblock {\em Phys. Rev. Lett.}, 89:130801--1, 2002.

\bibitem{Waters_magnetometer_NMR}
G.~S. Waters and P.~D. Francis.
\newblock {\em J. Sci. Instrum.}, 35:88, 1958.

\bibitem{Degen_sensor_review}
C.~L. Degen, F.~Reinhard, and P.~Cappellaro.
\newblock {\em Rev. Mod. Phys.}, 89:035002--1, 2017.

\bibitem{Taylor_diamond_magnetometer}
J.~M. Taylor, P.~Cappellaro, L.~Childress, L.~JIANG, D.~Budker, P.~R. Hemmer,
  A.~Yacoby, R.~Walsworth, and M.~D. Lukin.
\newblock {\em Nat. Phys.}, 4:810, 2008.

\bibitem{Wolf_NVensemble}
T.~Wolf, P.~Neumann, K.~Nakamura, H.~Sumiya, T.~Ohshima, J.~Isoya, and
  J.~Wrachtrup.
\newblock {\em Phys. Rev. X}, 5:041001--1, 2015.

\bibitem{Gruber_single_defect_spectroscopy}
A.~Gruber, A.~Dr{\"a}benstedt, C.~Tietz, L.~Fleury, J.~Wrachtrup, and C.~von
  Borczyskowski.
\newblock {\em Science}, 276:2012, 1997.

\bibitem{Dobrovitski_single_spin_control}
V.V. Dobrovitski, G.D. Fuchs, A.L. Falk, C.~Santori, and D.D. Awschalom.
\newblock {\em Annu. Rev. Condens. Matter Phys.}, 4:23, 2013.

\bibitem{Balasubramanian_nanoscale_imaging_magnetometry}
G.~Balasubramanian, I.Y. Chan, R.~Kolesov, M.~Al-Hmoud, J.~Tisler, C.~Shin,
  C.~Kim, A.~Wojcik, P.R. Hemmer, A.~Krueger, T.~Hanke, A.~Leitenstorfer,
  R.~Bratschitsch, F.~Jelezko, and J.~Wrachtrup.
\newblock {\em Nature}, 455:648, 2008.

\bibitem{Cole_decoherence_microscopy}
J.H. Cole and L.~C.~L. Hollenberg.
\newblock {\em Nanotechnology}, 20:1, 2009.

\bibitem{Dolde_electric_field_sensing_NV_spin}
F.~Dolde, H.~Fedder, M.W. Doherty, T.~N{\"o}bauer, F.~Rempp,
  G.~Balasubramanian, T.~Wolf, F.~Reinhard, L.~C.~L. Hollenberg, F.~Jelezko,
  and J.~Wrachtrup.
\newblock {\em Nature Physics}, 7:459, 2011.

\bibitem{Doherty_NV_centre_pressure_sensor}
M.~W. Doherty, V.V. Struzhkin, D.A. Simpson, L.P. McGuinness, Y.~Meng,
  A.~Stacey, T.J. Karle, R.J. Hemley, N.B. Manson, L.C.L. Hollenberg, and
  S.~Prawer.
\newblock {\em Phys. Rev. Lett.}, 112:047601--1, 2014.

\bibitem{Hodges_NV_centre_frequency_standard}
J.~S. Hodges, N.Y. Yao, D.~Maclaurin, C.~Rastogi, M.~D. Lukin, and D.~Englund.
\newblock {\em Phys. Rev. A}, 87:032118--1, 2013.

\bibitem{Kucsko_NV_centre_thermomtry}
G.~Kucsko, P.~C. Maurer, N.~Y. Yao, M.~Kubo, H.J. Noh, P.K. Lo, H.~Park, and
  M.D. Lukin.
\newblock {\em Nature}, 500:54, 2013.

\bibitem{Neumann_NV_centre_thermomtry}
P.~Neumann, I.~Jakobi, F.~Dolde, C.~Burk, R.~Reuter, G.~Waldherr, J.~Honert,
  T.~Wolf, A.~Brunner, J.H. Shim, D.~Suter, H.~Sumiya, J.~Isoya, and
  J.~Wrachtrup.
\newblock {\em Nano Lett.}, 13:2738, 2013.

\bibitem{Toyli_NV_centre_thermometry}
D.~M. Toyli, C.F. de~las Casasa, D.J. Christlea, V.V. Dobrovitski, and D.D.
  Awschalom.
\newblock {\em PNAS}, 110:8417, 2013.

\bibitem{Maiwald_enhanced_access_ion_trap}
R.~Maiwald, D.~Leibfried, J.~Britton, J.C. Bergquist, G.~Leuchs, and D.J.
  Wineland.
\newblock {\em Nature Physics}, 5:551, 2009.

\bibitem{Biercuk_trapped_ions_force_detection}
M.~J. Biercuk, H.~Uys†, J.W. Britton, A.P. VanDevender, and J.J. Bollinger.
\newblock {\em Nature Nanotecchnology}, 5:646, 2010.

\bibitem{Osterwalder_Rydberg_states_sensing}
A.~Osterwalders and F.~Merkt.
\newblock {\em Phys. Rev. Lett.}, 82:1831, 1999.

\bibitem{Braginsky_QND_measurements}
V.B. Braginsky, Y.I. Vorontsov, and K.S. Thorne.
\newblock {\em Science}, 209:547, 1980.

\bibitem{Levenson_QND_optics}
M.D. Levenson, R.M. Shelby, M.~Reid, and D.F. Walls.
\newblock {\em Phys. Rev. Lett.}, 57:2473, 1986.

\bibitem{Braginsky_QND}
V.B. Braginsky and F.Y. Khalili.
\newblock {\em Quantum {M}easurement, Ch. 4}.
\newblock Cambridge University Press, 1992.

\bibitem{Grangier_QND_meaurement_optics}
P.~Grangier, J.~Levenson, and J.~Poizat.
\newblock {\em Nature}, 396:537, 1998.

\bibitem{Nogues_QND_photon_measurement}
G.~Nogues, A.~Rauschenbeutel, S.~Osnaghi, M.~Brune, J.~M. Raimond, and
  S.~Haroche.
\newblock {\em Nature}, 400:239, 1999.

\bibitem{Gleyzes_quantum_jumps_of_light}
S.~Gleyzes, S.~Kuhr, C.~Guerlin, J.~Bernu, S.~Del{\'e}glise, U.B. Hoff,
  M.~Brune, J.M. Raimond, and S.~Haroche.
\newblock {\em Nature}, 446:297, 2007.

\bibitem{Hacker_NaturePhotonics_cat_states_Rb}
B.~Hacker, S.~Welte, S.~Daiss, A.~Shaukat, S.~Ritter, L.~Li, and G.~Rempe.
\newblock {\em Nature Photon}, 13:110, 2019.

\bibitem{Vlastakis_100Photon_cat_state}
B.~Vlastakis, G.~Kirchmair, Z.~Leghtas, S.E. Nigg, L.~Frunzio, S.M. Girvin,
  M.~Mirrahimi, M.H. Devoret, and R.J. Schoelkopf.
\newblock {\em Science}, 342:607, 2013.

\bibitem{Facon_SchrodingerCat_electrometer}
A.~Facon, E.K. Dietsche, D.~Grosso, S.~Haroche, J.M. Raimond, M.~Brune, and
  S.~Gleyzes.
\newblock {\em Nature}, 535:262, 2016.

\bibitem{Kitching_review}
J.~Kitching, S.~Knappe, and E.~A. Donley.
\newblock {\em IEEE Sensors Journal}, 11:1749, 2011.

\bibitem{Chen_spins_in_diamond_review}
M.~Chen, C.~Meng, Q.~Zhang, C.~Duan, F.~Shi, and J.~Du.
\newblock {\em Nat Sci Rev}, 5:346, 2018.

\bibitem{Lenz_magnetic_sensors_review}
J.~E. Lenz.
\newblock {\em Proceedings of The IEEE}, 78:973, 1990.

\bibitem{Josephson_superconductive_tunneling}
B.~D. Josephson.
\newblock {\em Phys. Lett.}, 1:251, 1962.

\bibitem{Anderson_tunneling_current_exp}
P.~W. Anderson and J.~M. Rowell.
\newblock {\em Phys. Rev. Lett.}, 10:230, 1963.

\bibitem{Jaclevic_currents_interference}
R.~C. Jaklevic, J.~Lambe, A.H. Silver, and J.E. Mercereau.
\newblock {\em Phys. Rev. Lett.}, 12:159, 1964.

\bibitem{Clarke_handbook}
J.~Clarke and A.~I. Braginski.
\newblock {\em The SQUID handbook: applications of {SQUID}s and {SQUID}
  systems}.
\newblock John {W}iley \& {S}ons, 2006.

\bibitem{Fagaly_SQUID_review}
R.~L. Fagaly.
\newblock {\em Rev. Sci. Instrum.}, 77:101101--1, 2006.

\bibitem{Clarke_highTc_SQUIDs}
J.~Clarke and R.~H. Koch.
\newblock {\em Science}, 242:217, 1988.

\bibitem{Vasyukov_nanoSQUID}
D.~Vasyukov, Y.~Anahory, L.~Embon, D.~Halbertal, J.~Cuppens, L.~Neeman,
  A.~Finkler, Y.~Segev, Y.~Myasoedov, M.L. Rappaport, M.E. Huber, and
  E.~Zeldov.
\newblock {\em Nature Nanotechnology}, 8:639, 2013.

\bibitem{Feynman_simulations_computer}
R.P. Feynman.
\newblock {\em Int J Theor Phys}, 21:467, 1982.

\bibitem{NISQ2018_Preskill}
J.~Preskill.
\newblock {\em Quantum}, 2:1, 2018.

\bibitem{Schneider_spectroscopy_sensing}
A.~Schneider, J.~Braumüller, L.~Guo, P.~Stehle, H.~Rotzinger, M.~Marthaler,
  A.V. Ustinov, and M.~Weides.
\newblock {\em Phys. Rev. A}, 97:062334--1, 2018.

\bibitem{Kristen_time-domain_sensing}
M.~Kristen, A.~Schneider, A.~Stehli, T.~Wolz, S.~Danilin, H.S. Ku, J.~Long,
  X.~Wu, R.~Lake, D.P. Pappas, A.V. Ustinov, and M.~Weides.
\newblock {\em npj Quantum Inf}, 6:1, 2020.

\bibitem{Honigl-Decrinis_power_sensor}
T.~H{\"o}nigl-Decrinis, R.~Shaikhaidarov, S.E. de~Graaf, V.N. Antonov, and O.V.
  Astafiev.
\newblock {\em Phys. Rev. Applied}, 13:024066--1, 2020.

\bibitem{Bylander_distortion_sensing}
J.~Bylander, M.S. Rudner, A.V. Shytov, S.O. Valenzuela, D.M. Berns, K.K.
  Berggren, L.S. Levitov, and W.D. Oliver.
\newblock {\em Phys. Rev. B}, 80:220506(R)--1, 2009.

\bibitem{Gustavsson_distortion_sensing}
S.~Gustavsson, O.~Zwier, J.~Bylander, Y.~Yan, F.~Yoshihara, Y.~Nakamura, T.P.
  Orlando, and W.D. Oliver.
\newblock {\em Phys. Rev. Lett.}, 110:040502--1, 2013.

\bibitem{Jerger_PRL}
M.~Jerger, A.~Kulikov, Z.~Vasselin, and A.~Fedorov.
\newblock {\em Phys. Rev. Lett.}, 123:150501--1, 2019.

\bibitem{Rol_APL}
M.A. Rol, L.~Ciorciaro, F.K. Malinowski, B.M. Tarasinski, R.E. Sagastizabal,
  C.C. Bultink, Y.~Salathe, N.~Haandbaek, J.~Sedivy, and L.~DiCarlo.
\newblock {\em Appl. Phys. Lett.}, 116:054001--1, 2020.

\bibitem{Larsen_PRL}
T.W. Larsen, K.D. Petersson, F.~Kuemmeth, T.S. Jespersen, P.~Krogstrup,
  J.~Nygård, and C.M. Marcus.
\newblock {\em Phys. Rev. Lett.}, 115:127001--1, 2015.

\bibitem{Lange_PRL}
G.~de~Lange, B.~van Heck, A.~Bruno, D.J. van Woerkom, A.~Geresdi, S.R.
  Plissard, E.P.A.M. Bakkers, A.R. Akhmerov, and L.~DiCarlo.
\newblock {\em Phys. Rev. Lett.}, 115:127002--1, 2015.

\bibitem{Giovannetti_1}
V.~Giovannetti, S.~Lloyd, and L.~Maccone.
\newblock {\em Nature Photon}, 5:222, 2011.

\bibitem{Giovannetti_4}
V.~Giovannetti, S.~Lloyd, and L.~Maccone.
\newblock {\em Phys. Rev. Lett.}, 96:010401--1, 2006.

\bibitem{Kitaev}
A.~Yu. Kitaev.
\newblock {\em arXiv:quant-ph/9511026}, 1995.

\bibitem{Quantum_Fourier}
W.~van Dam, G.M. D’Ariano, A.~Ekert, C.~Macchiavello, and M.~Mosca.
\newblock {\em Phys. Rev. Lett.}, 98:090501--1, 2007.

\bibitem{Danilin_magnetometry}
S.~Danilin, A.V. Lebedev, A.~Veps{\"a}l{\"a}inen, G.B. Lesovik, G.~Blatter, and
  G.S. Paraoanu.
\newblock {\em npj Quantum Inf}, 4:1, 2018.

\bibitem{Shlyakhov_metrology}
A.R. Shlyakhov, V.V. Zemlyanov, M.V. Suslov, A.V. Lebedev, G.S. Paraoanu, G.B.
  Lesovik, and G.~Blatter.
\newblock {\em Phys. Rev. A}, 97:022115--1, 2018.

\bibitem{Giovannetti_2}
V.~Giovannetti, S.~Lloyd, and L.~Maccone.
\newblock {\em Science}, 306:1330, 2004.

\bibitem{Giovannetti_3}
V.~Giovannetti, S.~Lloyd, and L.~Maccone.
\newblock {\em Phys. Rev. Lett.}, 108:260405--1, 2012.

\bibitem{Leibfried_Science}
D.~Leibfried, M.D. Barrett, T.~Schaetz, J.~Britton, J.~Chiaverini, W.M. Itano,
  J.D. Jost, C.~Langer, and D.J. Wineland.
\newblock {\em Science}, 304:1476, 2004.

\bibitem{Nagata_Science}
T.~Nagata, R.~Okamoto, J.L. O'Brien, K.~Sasaki, and S.~Takeuchi.
\newblock {\em Science}, 316:726, 2007.

\bibitem{Jones_Science}
J.A. Jones, S.D. Karlen, J.~Fitzsimons, A.~Ardavan, S.C. Benjamin, G.A.D.
  Briggs, and J.J.L. Morton.
\newblock {\em Science}, 324:1166, 2009.

\bibitem{Wang_NatCommun}
W.~Wang, Y.~Wu, Y.~Ma, W.~Cai, L.~Hu, X.~Mu, Y.~Xu, Z.J. Chen, H.~Wang, Y.P.
  Song, H.~Yuan, C.L. Zou, L.M. Duan, and L.~Sun.
\newblock {\em Nat Commun}, 10:1, 2019.

\bibitem{DiCarlo_cPhase}
L.~DiCarlo, J.M. Chow1, J.M. Gambetta, L.S. Bishop, B.R. Johnson, D.I.
  Schuster, JJ. Majer, A.~Blais, L.~Frunzio, S.M. Girvin, and R.J. Schoelkopf.
\newblock {\em Nature}, 460:240, 2009.

\bibitem{QuTiP}
J.R. Johansson, P.D. Nation, and F.~Nori.
\newblock {\em Comp. Phys. Comm.}, 184:1234, 2013.

\bibitem{Huelga_PRL}
S.F. Huelga, C.~Macchiavello, T.~Pellizzari, A.K. Ekert, M.B. Plenio, and J.I.
  Cirac.
\newblock {\em Phys. Rev. Lett.}, 79:3865, 1997.

\bibitem{Monz_PRL}
T.~Monz, P.~Schindler, J.T. Barreiro, M.~Chwalla, D.~Nigg, W.A. Coish,
  M.~Harlander, W.~H{\"a}nsel, M.~Hennrich, and R.~Blatt.
\newblock {\em Phys. Rev. Lett.}, 106:130506--1, 2011.

\bibitem{Harper_NatPhys}
R.~Harper, S.T. Flammia, and J.J. Wallman.
\newblock {\em Nat. Phys.}, 16:1184, 2020.

\bibitem{Lupke_PRX}
U.~von L{\"u}pke, F.~Beaudoin, L.M. Norris, Y.~Sung, R.~Winik, J.Y. Qiu,
  M.~Kjaergaard, D.~Kim, J.~Yoder, S.~Gustavsson, L.~Viola, and W.D. Oliver.
\newblock {\em PRX Quantum}, 1:010305--1, 2020.

\bibitem{Han_FundRes}
J.~Han, Z.~Li, J.~Zhang, H.~Xu, K.~Linghu, C.~Li, Y.and~Li, M.~Chen, Z.~Yang,
  J.~Wang, T.~Ma, G.~Xue, Y.~Jin, and H.~Yu.
\newblock {\em Fundamental Research}, 1:10, 2021.

\bibitem{Gebauer_AIP}
R.~Gebauer, N.~Karcher, D.~Gusenkova, M.~Spiecker, L.~Gr{\"u}nhaupt,
  I.~Takmakov, P.~Winkel, L.~Planat, N.~Roch, W.~Wernsdorfer, A.V. Ustinov,
  M.~Weber, M.~Weides, I.M. Pop, and O.~Sander.
\newblock {\em AIP Conference Proceedings}, 2241:020015--1, 2020.

\bibitem{all_microwave_gates}
J.M. Chow, A.D. C{\'o}rcoles, J.M. Gambetta, C.~Rigetti, B.R. Johnson, J.A.
  Smolin, J.R. Rozen, G.A. Keefe, M.B. Rothwell, M.B. Ketchen, and M.~Steffen.
\newblock {\em Phys. Rev. Lett.}, 107:080502--1, 2011.

\bibitem{QEC_theor_1}
P.W. Shor.
\newblock {\em Phys.Rev. A}, 52:R2493(R), 1995.

\bibitem{QEC_theor_2}
R.~Laflamme, C.~Miquel, J.P. Paz, and W.H. Zurek.
\newblock {\em Phys.Rev. Lett.}, 77:198, 1996.

\bibitem{QEC_exp_1}
M.~Reed, L.~DiCarlo, S.E. Nigg, L.~Sun, L.~Frunzio, S.M. Girvin, and R.J.
  Schoelkopf.
\newblock {\em Nature}, 482:382, 2012.

\bibitem{QEC_exp_2}
A.~C{\'o}rcoles, E.~Magesan, S.J. Srinivasan1, A.W. Cross, M.~Steffen, J.M.
  Gambetta, and J.M. Chow.
\newblock {\em Nat Commun}, 6:1, 2015.

\bibitem{QEC_exp_3}
Z.~Chen, K.J. Satzinger, J.~Atalaya, A.N. Korotkov, A.~Dunsworth, D.~Sank,
  C.~Quintana, M.~McEwen, R.~Barends, P.V. Klimov, S.~Hong, C.~Jones,
  P.~Petukhov, D.~Kafri, S.~Demura, B.~Burkett, C.~Gidney, A.G. Fowler,
  H.~Putterman, F.~Aleiner, I.and~Arute, K.~Arya, R.~Babbush, J.C. Bardin,
  A.~Bengtsson, A.~Bourassa, M.~Broughton, B.B. Buckley, D.A. Buell,
  N.~Bushnell, B.~Chiaro, R.~Collins, W.~Courtney, A.R. Derk, D.~Eppens,
  C.~Erickson, E.~Farhi, B.~Foxen, M.~Giustina, J.A. Gross, M.P. Harrigan, S.D.
  Harrington, J.~Hilton, A.~Ho, T.~Huang, W.J. Huggins, L.B. Ioffe, S.V.
  Isakov, E.~Jeffrey, Z.~Jiang, K.~Kechedzhi, S.~Kim, F.~Kostritsa,
  D.~Landhuis, P.~Laptev, E.~Lucero, O.~Martin, J.R. McClean, T.~McCourt,
  X.~Mi, K.C. Miao, M.~Mohseni, W.~Mruczkiewicz, J.~Mutus, O.~Naaman,
  M.~Neeley, M.~Neill, C.and~Newman, M.Y. Niu, T.E. O'Brien, A.~Opremcak,
  E.~Ostby, B.~Pat{\'o}, N.~Redd, P.~Roushan, N.C. Rubin, V.~Shvarts,
  D.~Strain, M.~Szalay, M.D. Trevithick, B.~Villalonga, T.~White, Z.J. Yao,
  P.~Yeh, A.~Zalcman, H.~Neven, S.~Boixo, YV. Smelyanskiy, Y.~Chen, A.~Megrant,
  and J.~Kelly.
\newblock {\em arXiv:2102.06132v1}, 2021.

\bibitem{Sekatski_theory}
P.~Sekatski, M.~Skotiniotis, J.~Ko{\l}ody{\'n}ski, and W.~D{\"u}r.
\newblock {\em Quantum}, 1:1, 2017.

\bibitem{HL_in_metrology}
S.~Zhou, M.~Zhang, J.~Preskill, and L.~Jiang.
\newblock {\em Nat Commun}, 9:1, 2018.

\bibitem{QEC_against_relaxation}
D.A. Herrera-Mart{\'i}, T.~Gefen, D.~Aharonov, N.~Katz, and A.~Retzker.
\newblock {\em Phys.Rev. Lett.}, 115:200501--1, 2015.

\end{thebibliography}
\bibliographystyle{unsrt}

\end{document}